\documentclass[12pt]{article}
\usepackage{amsmath}
\usepackage{graphicx}
\usepackage{enumerate}
\usepackage{natbib}
\usepackage[margin=1.5in]{geometry}
\usepackage{amsmath}%
\usepackage{amsfonts}%
\usepackage{amssymb}%
\usepackage{graphicx}
\usepackage{natbib}
\usepackage{booktabs}
\usepackage{url}
\usepackage{listings}
\usepackage{xcolor}
\usepackage{fancyvrb}
\input{preambleKnit.tex}
\newcommand{\bse}{\boldsymbol{\eta}}
\newcommand{\bsZ}{\boldsymbol{Z}}
\newcommand{\bsG}{\boldsymbol{G}}

\newcommand{\bds}[1]{\boldsymbol{#1}}

\newcommand{\blind}{0}

\begin{document}

\def\spacingset#1{\renewcommand{\baselinestretch}%
{#1}\small\normalsize} \spacingset{1}

\if0\blind
{
  \title{\bf  Avoiding Bias Due to Nonrandom Scheduling When Modeling Trends in Home-Field Advantage}
  \author{Andrew Karl}
  \date{October 22, 2026}
  \maketitle
} \fi

\if1\blind
{
  \bigskip
  \bigskip
  \bigskip
  \begin{center}
    {\large\bf  Avoiding Bias Due to Nonrandom Scheduling When Modeling Trends in Home-Field Advantage}
\end{center}
  \medskip
} \fi

\bigskip
\begin{abstract}
Existing approaches for estimating home-field advantage (HFA) include modeling the difference between home and away scores as a function of the difference between home and away team ratings that are treated either as fixed or random effects. We uncover an upward bias in the mixed model HFA estimates that can arise when the schedule -- and thus the random effect design matrix -- depends on team strengths, and explore why the fixed effects model is not subject to the same bias. Intraconference HFAs and standard errors are calculated for each of 3 college sports and 3 professional sports over the 2000--2017 seasons and then fitted with conference-specific slopes and intercepts to measure the potential linear population trend in HFA.
\end{abstract}

\noindent%
\vfill

\newpage
\spacingset{1.45} % DON'T change the spacing!
\section*{Revision note}
The general theory of mixed model bias and diagnostics was subsequently developed by \citet{KarlZimmerman2021}. This revision retains the original 2000--2017 sports analysis, replaces the exploratory diagnostic argument with a direct bias calculation, and corrects the women's college basketball Phase II fit by constraining its random effect covariance matrix to be positive semidefinite. The substantive conclusions are unchanged.
\section{Introduction}
\label{sec:intro}

There has long been observed a ``home-field advantage'' across a variety of sports \citep{harville77,harville94,harville03,pollard05,marshall,wang,goumas}. This advantage may be defined as ``the net effect of several factors that generally benefit the home team and disadvantage the visiting team'' \citep{harville94}.

Recently, sports journalists and researchers have taken interest in a perceived decline in the home-field advantage (HFA) in college football and basketball. \cite{omaha} notes the decrease in college football HFA by looking at win proportions for the home team within conferences. \cite{si2} makes a similar note of declining proportion of wins for home teams in intraconference NCAA Division I men's basketball games. However, these analyses assume an underlying binomial model for the number of games in a season won by home teams \citep{pollard05} and that game outcomes are independent with an equal probability of the home team winning. Simply tracking the proportion of games won by the home team does not control for the variability in team abilities, nor consider the potential biasing impact of nonrandom scheduling on the estimate, nor provide accurate standard errors for the proportions.

HFA can be approached from two different perspectives:
\begin{enumerate}
\item{\textbf{scoring HFA:} the difference in the expected scores of the home and away teams within a game, after the  strengths of each team have been accounted for.}
\item{\textbf{win propensity HFA:} the increased win propensity of a team when playing an opponent at home, beyond the predicted propensity of that team defeating the same opponent at a neutral site.} 
\end{enumerate}

This investigation studies whether or not there is a linear trend in the scoring HFA of 6 sports over the 18 seasons from 2000 through 2017: Men's and Women's NCAA Division I basketball, the NCAA Football Bowl Subdivision (FBS), the National Football League (NFL), the National Basketball Association (NBA), and the Women's National Basketball Association (WNBA). The teams of each of these sports are divided into conferences, which are clusters of teams that play each other with higher probability than would be expected if the schedules were randomized across the sport. We are interested in knowing if trends in scoring HFA differ across the conferences.

In the first phase of this investigation, we present a model for estimating HFA (independently) within each conference and in each season. \citet{pollard05} acknowledge that schedules may be unbalanced, but do not explore the potential impact on HFA estimates. However, \citet[Section 5]{harville77} notes that the estimate for HFA resulting from the mean of the differences between home and away scores is ```biased' upwards because, for financial reasons, the better teams are more generally able to arrange schedules with more games at home than away.'' However, we will see that a simplified version of the mixed model estimate for HFA proposed by \citet{harville77} is itself subject to upward bias in this situation, by examining a fixed and a random effect model for scoring HFA. 

The second phase of the analysis fits a potential linear trend in the yearly conference estimates of HFA that were produced in Phase I. In the college sports, which typically have at least 10 conferences, this phase implements a weighted random coefficient model using the Phase I intraconference HFA estimates as the response and the reciprocal of the squared standard errors from Phase I as weights. The model allows for correlated conference-specific random intercepts and slopes. Due to the presence of only two conferences in each professional sport, the conference-specific intercepts and slopes are instead treated as fixed effects for these sports.

\section{Methods}\label{sec:methods}

\subsection{Phase I: Estimating HFA}

This section presents both a fixed effects model and a mixed effects model for the home team margins of victory, $d_i=y_{H_i}-y_{A_i}$, where $y_{H_i}$ and $y_{A_i}$ are the home and away team scores, respectively, for game $i=1,...,n$. With $N$ teams in a data set, the pattern of game opponents and locations (the schedule) is recorded in an $n\times N$ matrix $\bds{Z}$ as follows: if team $T_H$ hosted team $T_A$ in game $i$, then the $i$-th row of $\bds{Z}$ consists of all zeros except for a $1$ in the column corresponding to $T_H$ and a $-1$ in the column corresponding to $T_A$.

\subsubsection{Fixed Effects Model for Scoring Advantage}
The first plausible model for scoring HFA fits a fixed effect $\lambda$ for the HFA and vector of fixed team effects $\bds{\beta}=\left(\beta_1,\ldots,\beta_N\right)^{\prime}$ where the difference between team effects provides an estimate of the score difference in a game between the two teams on a neutral field. For each game $i$, this model assumes
\begin{equation}
d_i\mid\bsZ\sim N\left(\lambda + \beta_{H_i} - \beta_{A_i}, \sigma^2 \right)
\end{equation}
where $H_i$ and $A_i$ are the indices for the home and away teams, respectively, in game $i$. The residuals are assumed to be independent, leading to an overall model 
\begin{equation}\label{eq:femodel}
\bds{d}=\lambda*\bds{1} + \bds{Z}\bds{\beta} + \bds{\epsilon}
\end{equation}
where $\bds{1}$ is a vector of 1's, $\bds{\epsilon}\sim N(\bds{0},\sigma^2\bds{I})$ where $\bds{I}$ is the identity matrix, and $\bds{d}=(d_1,\ldots,d_n)^{\prime}$. This model appears previously as Model 2 of \cite{harville94} and Model 1 of \cite{harville03}.

The fixed effect model matrix $\bds{Z}$  is not full rank and the individual team effects $\beta_i$ are not estimable. However, $\lambda$ is estimable, as are the pairwise differences between team effects, provided there is sufficient mixing of the teams \citep[Section 5.2.1]{stroup}. For example, in a conference of three teams (A, B, C), $\lambda$ will be estimable if any two teams have a home and home series, or if, for example, team B plays at team A, team A plays at team C, and team C plays at team B. Provided this is the case  and letting $M^+$ represent the Moore-Penrose inverse of a matrix $M$, solutions to model (\ref{eq:femodel}) are given by 
\begin{equation}\label{eq:febeta}
\left[
\begin{array}{c}
\widehat{\lambda}\\
\bds{\widehat{\beta}}
\end{array}
\right]
=\left[
\begin{array}{c|c}
\left(\bds{1^{\prime}}\bds{1}\right)&\left(\bds{1^{\prime}}\bsZ\right)\\
\hline
\left(\bsZ^{\prime}\bds{1}\right)&\left(\bsZ^{\prime}\bsZ\right)
\end{array}
\right]^{+}
\left[
\begin{array}{c}
\bds{1^{\prime}}\\
\hline
\bsZ^{\prime}
\end{array}
\right]
\bds{d}
\end{equation}
 
The schedule $\bsZ$ may itself be stochastic and depend on team strengths, in the sense that better teams may choose to play weaker opponents or more games at home than away. Provided that $\lambda$ is estimable and $\mathrm{E}(\bds{\epsilon}\mid\bsZ)=\bds{0}$, taking the conditional expectation of (\ref{eq:febeta}) shows that the fixed effects estimate of $\lambda$ is unbiased conditional on the schedule. Independence of $\bsZ$ and $\bds{\epsilon}$ is sufficient for this mean condition. If $\bds{1^{\prime}}\bsZ=\bds{0}$, each team plays equally many home and away games and the estimated HFA simplifies to $\widehat{\lambda}=\overline{\bds{d}}$.

\subsubsection{Mixed Model for Scoring Advantage}
Alternatively, the teams may be modeled with random effects $\bds{\eta}=\left(\eta_1,\ldots,\eta_N\right)^{\prime}\sim N\left(\bds{0},\sigma^2_g\bds{I}\right)$ that are assumed to be independent of the error terms. The standard mixed model also assumes independence of $\bsZ$ and $\bse$; below we examine the consequences of relaxing this assumption. The remaining details are the same as model (\ref{eq:femodel}), with difference in home and away scores $d_i$ being modeled conditional on the random team effects and schedule as
\begin{equation}
d_i\mid\bse,\bsZ\sim N\left(\lambda + \eta_{H_i} - \eta_{A_i}, \sigma^2 \right)
\end{equation}
producing an overall model
\begin{equation}\label{eq:mixed}
\bds{d}\mid\bse,\bsZ\sim N(\lambda*\bds{1} + \bds{Z}\bds{\eta},\sigma^2\bds{I})
\end{equation}
\citet[Equation 2.1]{harville77} considers a generalization of this model for ranking sports teams. An advantage of placing a distributional assumption on the team effects is that it provides a form of regularization for the model and avoids the estimability concern for the team effects.

As with the fixed effect model (\ref{eq:febeta}), the mixed model estimate for $\lambda$ is equal to $\overline{\bds{d}}$ when $\bds{1^{\prime}}\bsZ=\bds{0}$ \citep[Equation 5]{henderson75}. Thus the two estimates agree for a balanced schedule, even if the schedule depends on team strengths. For an unbalanced schedule, the source of possible bias can be stated directly. First hold the variance components fixed, with $\sigma^2>0$ and $\sigma_g^2\geq0$, and write
\[
\bds{V}=\sigma_g^2\bsZ\bsZ^{\prime}+\sigma^2\bds{I},
\qquad
\bds{a}(\bsZ)=\frac{\bds{V}^{-1}\bds{1}}
{\bds{1^{\prime}}\bds{V}^{-1}\bds{1}}.
\]
The generalized least squares estimate is
\begin{equation}\label{eq:gls}
\widetilde\lambda=\bds{a}(\bsZ)^{\prime}\bds{d}.
\end{equation}
Since $\bds{a}(\bsZ)^{\prime}\bds{1}=1$, if $\mathrm{E}(\bds{\epsilon}\mid\bsZ,\bse)=\bds{0}$ and the expectations exist, then
\begin{equation}\label{eq:glsbias}
\mathrm{E}(\widetilde\lambda-\lambda\mid\bsZ)
=\bds{a}(\bsZ)^{\prime}\bsZ\mathrm{E}(\bse\mid\bsZ).
\end{equation}
The marginal bias is therefore $\mathrm{E}[\bds{a}(\bsZ)^{\prime}\bsZ\bse]$. The condition $\mathrm{E}(\bse\mid\bsZ)=\bds{0}$ gives zero bias in this calculation and is satisfied when the full matrix $\bsZ$ and $\bse$ are independent with $\mathrm{E}(\bse)=\bds{0}$. Dependence can produce nonzero bias, but need not do so. Normality is not needed for (\ref{eq:glsbias}). When $\bds{1^{\prime}}\bsZ=\bds{0}$, $\bds{a}(\bsZ)=\bds{1}/n$ and $\bds{a}(\bsZ)^{\prime}\bsZ=\bds{0}$; this simplification also holds with fitted variance components. The fitted models below use REML variance estimates; away from this balanced case their weights also depend on the data, so (\ref{eq:glsbias}) describes the fixed-variance mechanism rather than the exact bias of the fitted estimator. The simulations below examine the behavior of those fitted models. A more general treatment and a developed diagnostic are given by \citet{KarlZimmerman2021}.

\subsection{Phase II: Fitting a Trend in HFA}\label{sec:phase2}
Once the estimated intraconference HFAs and associated standard errors have been obtained from either model (\ref{eq:femodel}) or model (\ref{eq:mixed}), we will check for the presence of a population-wide linear trend in HFA within each sport. In the application in Section~\ref{sec:results}, separate Phase I models are fit for the different conferences within sports in each year. There are many such conferences for the college sports, but only two for each of the professional sports. 
\subsubsection{College Sports}\label{sec:collegeII}
For the college sports, we fit a model with a linear trend in time and correlated random slopes and intercepts for conferences. Letting $\lambda_{ij}$ represent the estimated HFA for conference $j$ in year $i$, we fit
\begin{equation}\label{eq:rancoef}
\lambda_{ij} = \left(\alpha_0 + b_{0j}\right) + \left(\alpha_1 + b_{1j}\right)t_i+\epsilon_{ij}
\end{equation}
where $t_i$ is the year (indexed with 2017 as time 0), $\alpha_0$ and $\alpha_1$ are fixed effects for the population intercept and slope, respectively, and the errors are independent and distributed $\epsilon_{ij}\sim N \left(0,{\sigma^2_{\lambda}}{w_{ij}^2}\right)$ where $w_{ij}$ is the standard error of the HFA for conference $j$ in year $i$ from Phase I: this weighting incorporates information about the accuracy of the point estimates from Phase I. The conference-level random intercepts and slopes are assumed to be independent across conferences, and independent of $\epsilon$,  distributed as $\left(b_{0j},b_{1j}\right)\sim N\left(\bds{0},\bsG\right)$ where $\bsG$ is a covariance matrix with lower triangle $\left(\sigma^2_1,\sigma_{12},\sigma^2_2\right)^{\prime}$.

Testing for the significance of the contributions of the conference-level random slopes and intercepts is not amenable to the standard likelihood ratio test since the null hypothesis of a zero variance component lies on the boundary of the parameter space \citep{scheipl}: a simulation is used to assess the magnitude of the observed restricted loglikelihood statistic under the null hypothesis $\bsG=\bds{0}$. Random effects are retained in the model regardless of this test, to allow for conference-level heterogeneity. The resulting fixed effect estimate $\widehat{\alpha}_1$ represents sport-wide linear trend in HFA, while $\widehat{\alpha}_0$ estimates the population HFA in 2017.
\subsubsection{Professional Sports}\label{sec:proII}
Each of the considered professional sports are split into two conferences, which will be denoted A and B. In the NBA and WNBA, these are the Western and Eastern conferences, respectively. In the NFL, they are the NFC and the AFC, respectively. With only two conferences, the HFA estimates from each conference are not able to support the estimation of $\bsG$ in (\ref{eq:rancoef}). Instead, the conference intercepts and slopes are fit with fixed effects in model (\ref{eq:fullfixed}) where the errors are weighted and distributed as described in Section~\ref{sec:collegeII}.  The full model is compared via a likelihood ratio test (using maximum likelihood estimates) to model (\ref{eq:fixed1}), which assumes identical slopes and intercepts for both conferences, and to model (\ref{eq:fixed2}), which drops the linear trend in time from the full model. 
\begin{align}
\lambda_{ij} =& \beta_{0A} + \beta_{1A}t_i +\left(\beta_{0B}+\beta_{1B}t_i\right)I\left(j=B\right) +\epsilon_{ij}\label{eq:fullfixed}\\
\lambda_{ij} =& \beta_{0A} + \beta_{1A}t_i +\epsilon_{ij}\label{eq:fixed1}\\
\lambda_{ij} =& \beta_{0A}  +\beta_{0B}I\left(j=B\right) +\epsilon_{ij}\label{eq:fixed2}
\end{align}
In the full model, $\beta_{0A}$ and $\beta_{1A}$ are the intercept and slope for conference A, while $\beta_{0B}$ and $\beta_{1B}$ are the corresponding B-minus-A contrasts. In model (\ref{eq:fixed1}), the intercept and slope are common to both conferences.
\subsection{Estimation}\label{sec:estimation}

The Phase I restricted maximum likelihood (REML) estimates for the linear mixed model (\ref{eq:mixed}) are obtained with an EM algorithm \citep{laird82,pxem} via the R package \texttt{mvglmmRank} \citep{mvglmmRank}, which also uses the resulting model matrices to construct the fixed effect estimates (\ref{eq:febeta}). The Phase II models of Sections~\ref{sec:collegeII} and \ref{sec:proII} were originally fit in SAS PROC MIXED. In this revision, the women's college basketball Phase II model is refit by REML with a positive semidefinite covariance constraint using \texttt{lme4} 1.1-37; its fixed effect confidence intervals and tests use the Kenward--Roger approximation in \texttt{lmerTest} 3.2-1 and \texttt{pbkrtest} 0.5.5. The original unconstrained fit returned an inadmissible random effect covariance matrix. The other fitted models are retained. The original analysis code and result files, together with the corrected Phase II fit and its inputs, are supplied as ancillary material with this revision. The source code can be adapted to perform similar analyses on results from other sports using the database provided by \citet{masseyr}.

\section{Application to College and Professional Basketball and Football}\label{sec:results}

\subsection{Data}

Scores from the 2000--2017 seasons for each of the six sports of Men's and Women's NCAA Division I basketball, NCAA Football Bowl Subdivision (FBS), NFL, the WNBA, and the NBA are furnished by the website of \citet{masseyr}, with some missing WNBA results obtained from WNBA.com. In all cases, neutral site games are excluded, and college results are limited to intradivision games between Division I teams for basketball and FBS teams for football. The choice of 2000 as a starting year is further explained in the appendix.

\subsection{Phase I: Estimating the HFA}

\subsubsection{Application to Full Season Data}

\begin{figure}
\caption{Fixed effect model estimates (dotted) of the scoring HFA are plotted with corresponding mixed model (dashed) estimates and the mean of $\bds{d}$ (solid) across seasons of NCAA Men's Basketball (left) and the NBA (right). The lines in the NBA plot overlap.  }
\label{plot:marg1}
\centering
\includegraphics[scale=0.5]{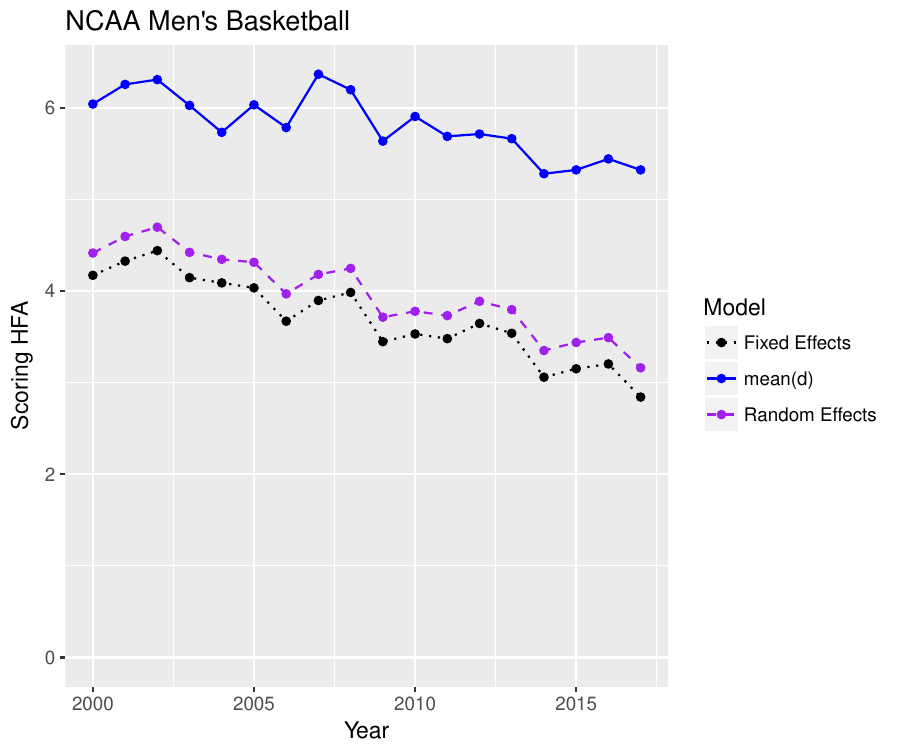}
\includegraphics[scale=0.5]{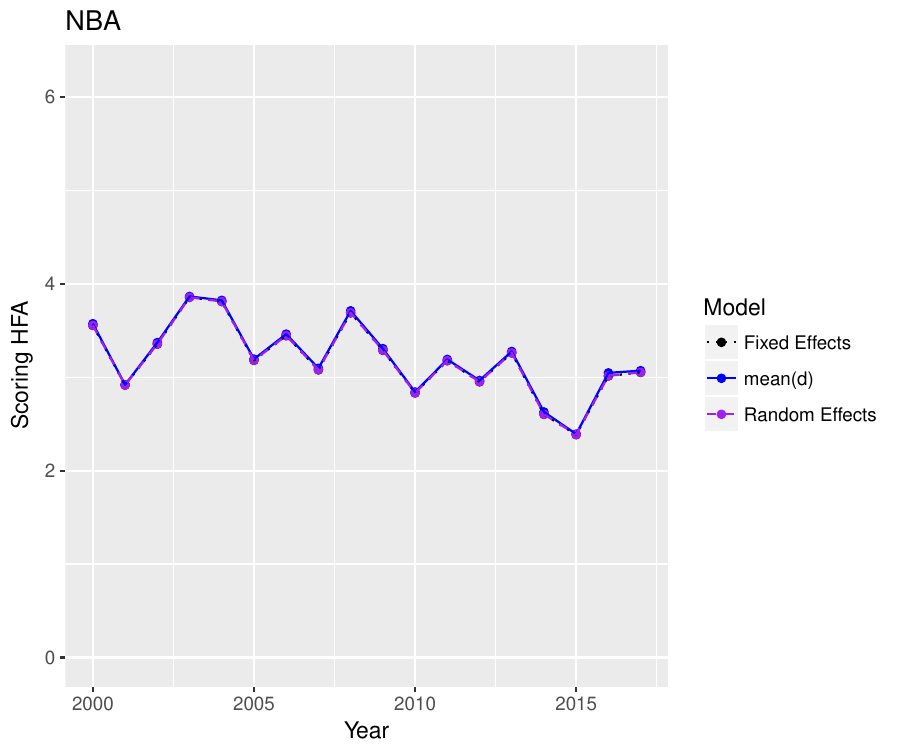}
\end{figure}

In an initial application to full seasons of NCAA men's basketball results (ignoring conference divisions and thus using every game played during each season), Figure~\ref{plot:marg1} shows that the mean $\overline{\bds{d}}$ is uniformly larger than the estimated HFA from the mixed model (\ref{eq:mixed}), which in turn is uniformly larger than the estimates from the fixed effects model (\ref{eq:femodel}). However, HFA estimates from these three approaches are identical to each other for the same NBA seasons. The same pattern appears in the other college and professional sports, with three distinct estimates in each of the college sports and three nearly identical estimates in each season of each of the professional sports. A bootstrap simulation explores whether these differences are associated with the alignment between team strengths and the schedule.

\subsubsection{Simulation for Full Seasons}\label{sec:fullsim}
To assess the behavior of the HFA estimates produced by the fixed and mixed effect models given the potentially nonrandom schedules, $\bsZ$, we run two simulations in each full season of each sport (ignoring conferences) using the fitted team effects $\widehat{\bse}$ and error terms $\widehat{\bds{e}}$ from the mixed model (\ref{eq:mixed}). The first simulation generates observations with a HFA of 3 points by resampling the residuals with replacement (bootstrapping) as defined in (\ref{eq:ysim1}). In the second case, presented by (\ref{eq:ysim2}), the team effects are also resampled (without replacement).

The first simulation represents a scenario in which the season is played repeatedly according to the same schedule, while the second represents a scenario in which the teams are shuffled before assigning them to the same schedule structure, $\bsZ$. Letting $\bds{s_1}(\bds{x})$ represent a function that samples the elements of $\bds{x}$ with replacement (or without replacement for $\bds{s_0}$),
\begin{align}
\bds{y}_{sim1}&=3*\bds{1}+\bds{Z}\bds{\widehat{\eta}}+\bds{s_1}\left(\bds{\widehat{e}}\right)\label{eq:ysim1}\\
\bds{y}_{sim2}&=3*\bds{1}+\bds{Z}\bds{s_0}\left(\bds{\widehat{\eta}}\right)+\bds{s_1}\left(\bds{\widehat{e}}\right)\label{eq:ysim2}
\end{align}
2000 replicates of $\bds{y}_{sim1}$ and $\bds{y}_{sim2}$ are generated for each season of each sport. These simulated score vectors are fit with both the fixed and mixed effect models. For each year $i$, the mean, $m_i$, of the 2000 estimates of $\lambda$, and the coverage probability of the 95\% confidence intervals is recorded. The means of these values over all 18 seasons is then reported in Table~\ref{tab:ysim1}, along with the p-value from the two-sided $t$-test of the null hypothesis that the $m_i$ were drawn from a population with mean 3.

Both models recover values close to the simulated HFA of 3 points in all sports when fitting the simulation with randomized team assignments, $\bds{y}_{sim2}$. Likewise, both models show only small departures from 3 in the professional sports, which have more balanced schedules than the college sports, when the schedule remains fixed in its current configuration in $\bds{y}_{sim1}$.

However, only the fixed effect model consistently recovers values close to 3 for $\bds{y}_{sim1}$ in the college sports: the mixed effect model shows upward bias in its estimate of the college HFAs along with poor coverage probabilities and extremely small p-values. The standard mixed model (\ref{eq:mixed}) assumes independence between team effects and the schedule. This assumption is questionable when stronger college teams are able to schedule more home than away games. The use of random effects to condition on team abilities ameliorates, but does not eliminate, the bias associated with these nonrandom assignments, as illustrated in Figure~\ref{plot:marg1} and Table~\ref{tab:ysim1}.

\begin{table}
\caption{Results for fitting 2000 simulations of $y_{sim1}$ and $y_{sim2}$  in each of the eighteen seasons for each sport, with $\lambda=3.0$.  Coverage probabilities are for 95\% confidence intervals. The p-value refers to the two-sided $t$-test of the null hypothesis that the means of the 2000 simulations in each year are drawn from a population with a mean of 3. CFB represents college football, while CBB-M and CBB-W represent men's and women's college basketball, respectively.}
\label{tab:ysim1}
\vspace{0.5em}

\begin{tabular}{lrrrrrr}
  \toprule
 & NFL & NBA & WNBA & CFB & CBB-M & CBB-W \\ 
  \midrule
\multicolumn{6}{l}{\textit{Fixed team effects for} $y_{sim1}$}\\
$\text{mean}(\widehat{\lambda})$ & 3.00 & 3.00 & 3.00 & 3.00 & 3.00 & 3.00 \\ 
  Coverage Prob. & 0.95 & 0.95 & 0.95 & 0.95 & 0.95 & 0.95 \\ 
  p-value & 0.40 & 0.77 & 0.63 & 0.66 & 5e-03 & 0.87 \\ 
   \midrule
\multicolumn{6}{l}{\textit{Random team effects for} $y_{sim1}$}\\
$\text{mean}(\widehat{\lambda})$ & 3.02 & 3.00 & 3.00 & 3.37 & 3.26 & 3.13 \\ 
  Coverage Prob. & 0.95 & 0.95 & 0.95 & 0.89 & 0.62 & 0.87 \\ 
  p-value& 0.02 & 0.37 & 0.12 & 4e-15 & 1e-19 & 4e-18 \\ 
   \midrule
\multicolumn{6}{l}{\textit{Fixed team effects for} $y_{sim2}$}\\
$\text{mean}(\widehat{\lambda})$ & 3.00 & 3.00 & 3.01 & 2.99 & 3.00 & 3.00 \\ 
  Coverage Prob. & 0.95 & 0.95 & 0.95 & 0.95 & 0.95 & 0.95 \\ 
  p-value & 0.65 & 0.56 & 0.03 & 0.02 & 0.21 & 0.47 \\ 
   \midrule
\multicolumn{6}{l}{\textit{Random team effects for} $y_{sim2}$}\\
$\text{mean}(\widehat{\lambda})$ & 3.00 & 3.00 & 3.01 & 2.99 & 3.00 & 3.00 \\ 
  Coverage Prob. & 0.95 & 0.95 & 0.95 & 0.95 & 0.95 & 0.95 \\ 
  p-value& 0.66 & 0.57 & 0.03 & 0.02 & 0.29 & 0.41 \\ 
   \bottomrule
\multicolumn{6}{l}{}\\
\end{tabular}
\end{table}

\begin{table}
\caption{Results for fitting 500 simulations of $y_{sim1}$ in each conference in each of the eighteen seasons, for each sport, with $\lambda=3.0$.  Coverage probabilities are for 95\% confidence intervals. The p-value refers to the two-sided $t$-test of the null hypothesis that the population of simulated conference  means (averages over the 500 simulations in each conference in each season)  has mean equal to 3. CFB represents college football, while CBB-M and CBB-W represent men's and women's college basketball, respectively.}
\label{tab:conf7}
\vspace{0.5em}
\begin{tabular}{lrrrrrr}
  \toprule
 & NFL & NBA & WNBA & CFB & CBB-M & CBB-W \\ 
  \midrule
\multicolumn{6}{l}{\textit{Fixed team effects for} $y_{sim1}$}\\
$\text{mean}(\widehat{\lambda})$ & 2.99 & 3.00 & 3.00 & 3.00 & 3.00 & 3.00 \\ 
  Coverage Prob. & 0.95 & 0.95 & 0.95 & 0.94 & 0.95 & 0.95 \\ 
  p-value & 0.38 & 0.76 & 0.58 & 0.51 & 0.99 & 0.35 \\ 
   \midrule
\multicolumn{6}{l}{\textit{Random team effects for} $y_{sim1}$}\\
$\text{mean}(\widehat{\lambda})$ & 3.02 & 3.00 & 3.01 & 2.99 & 3.01 & 3.03 \\ 
  Coverage Prob. & 0.95 & 0.95 & 0.95 & 0.94 & 0.94 & 0.95 \\ 
  p-value& 0.09 & 0.39 & 0.12 & 0.48 &6e-08 & 8e-20 \\ 
   \bottomrule
   \end{tabular}
\end{table}

\subsubsection{Restriction to Intraconference Games}
Games played within conferences of these sports tend to be more balanced (in terms of opponents' strength and number of home and away games) than games played between conferences. Instead of estimating HFA using full season results, we could also fit each set of intraconference games separately within each season: interconference game results are discarded. The simulations below assess whether this restriction reduces the bias associated with nonrandom scheduling.

The simulation (\ref{eq:ysim1}) was run with 500 iterations for each of the intraconference schedules in each season of each sport (Table~\ref{tab:conf7}). While the restriction to intraconference games largely eliminated the bias in the mixed model estimate of $\lambda$, there is still some remaining evidence of upward bias in these estimates. For example, in women's college basketball the mixed model for the full season simulations produced a mean estimate of $3.13$ with a coverage probability of 0.87. This improved to an estimate of $3.03$ and a coverage probability of 0.95 when restricting to intraconference games; however, the p-value of 8e-20 from the two-sided $t$-test indicates that  upward bias is still a systematic problem for the random team effect model. As a result, the Phase II models presented in Section~\ref{sec:phase2} will make use of the HFA estimates from the fixed team effect model (\ref{eq:femodel}).

An additional benefit of restricting to intraconference games is the potential for a comparison of HFA trends within conferences across seasons. Of course, this limits the inference space to intraconference games, requiring either  additional assumptions or additional modeling structure to generalize conclusions to interconference games.

\subsection{Phase II: Trends in Home-Field Advantage}
\label{ssec:hfe}

\begin{figure}
\caption{Each point represents the estimated scoring HFA for a conference within each season and sport. Graphs have common limits for the y-axis. In the college sports, the lines correspond to different conferences. In the professional sports, the triangles (circles) and solid (dashed) lines represent the NFC (AFC) conference in the NFL and the Western (Eastern) Conference in the NBA and WNBA.}
\label{plot:HFE_b}
\centering
\includegraphics[scale=0.5]{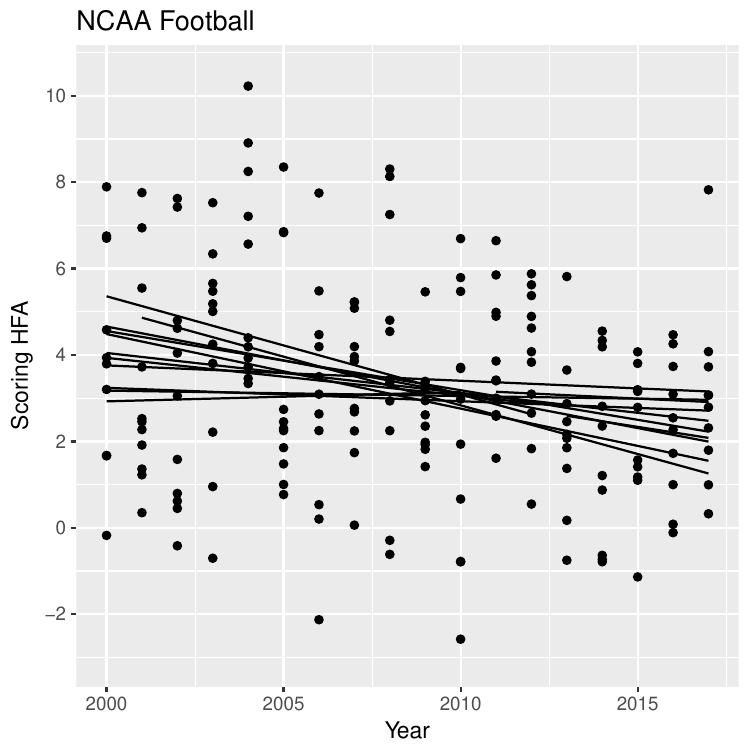}
\includegraphics[scale=0.5]{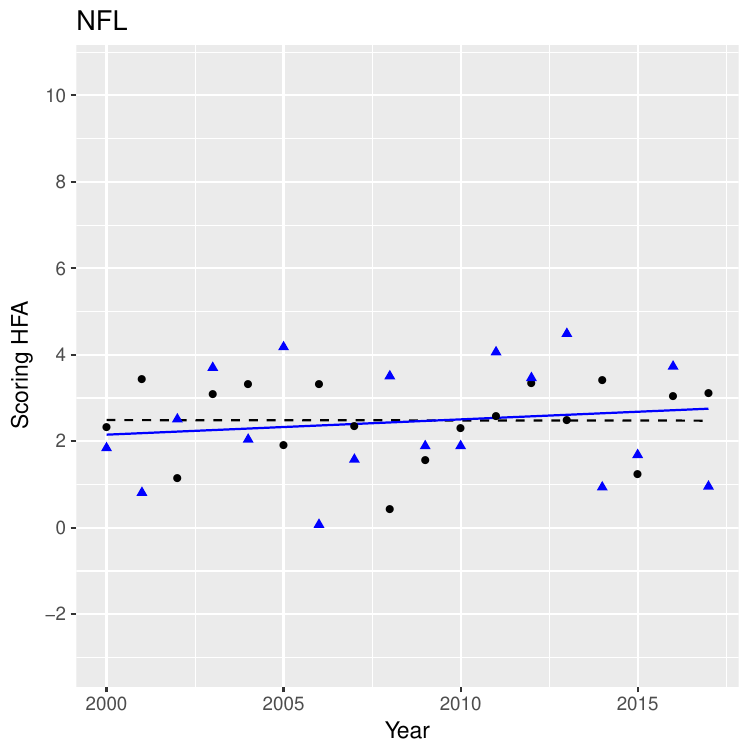}
\includegraphics[scale=0.5]{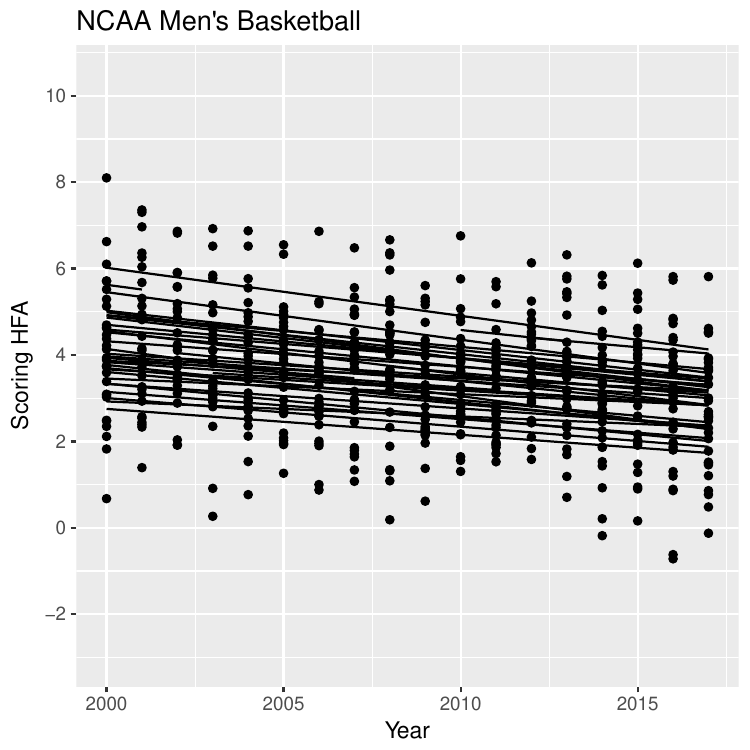}
\includegraphics[scale=0.5]{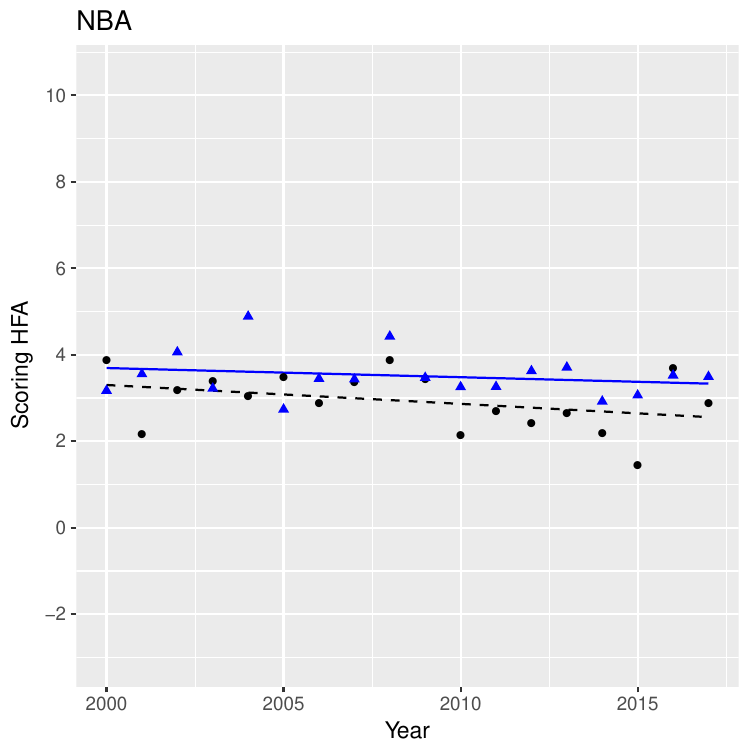}
\includegraphics[scale=0.5]{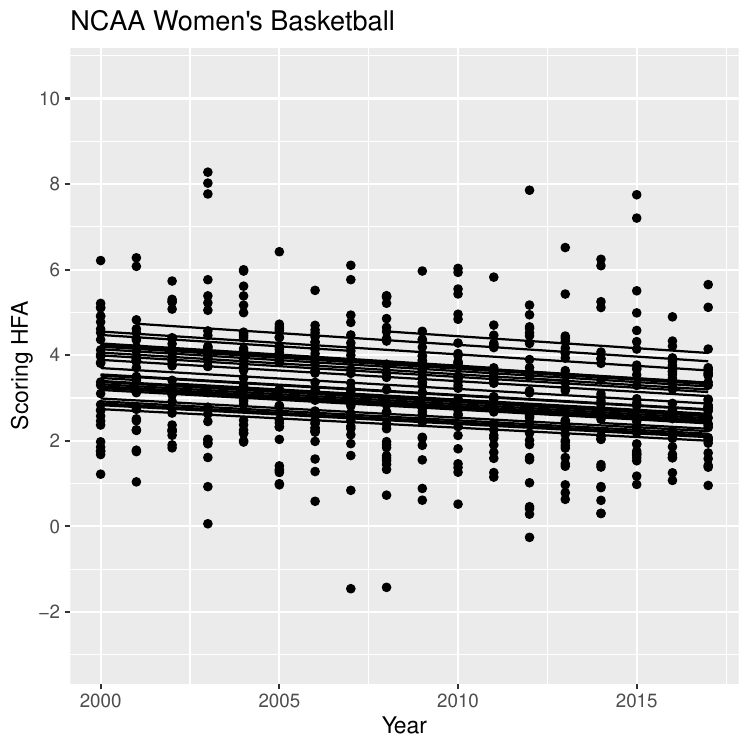}
\includegraphics[scale=0.5]{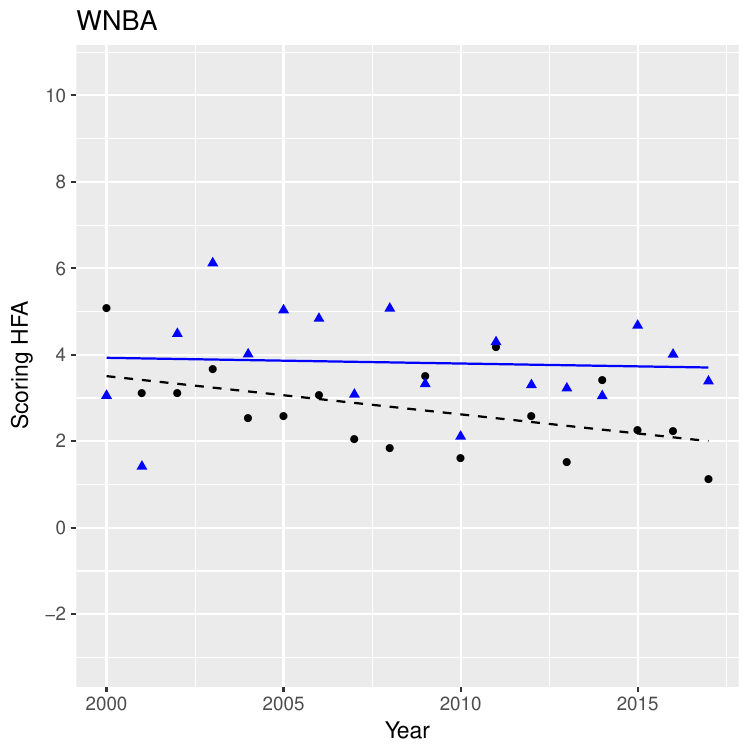}
\end{figure}

After fitting separate HFA estimates for each conference in each season of each sport, the resulting point estimates and standard errors are used to fit the Phase II models of Sections~\ref{sec:collegeII} and \ref{sec:proII}. The results for the college sports appear in Table~\ref{tab:tabone} and those for the professional sports appear in Table~\ref{tab:pro}. The conference-specific trends in scoring HFA are plotted in Figure~\ref{plot:HFE_b}.

\begin{table}
\caption{Phase II results of fitting the random coefficient model (\ref{eq:rancoef}) to the Phase I estimated HFAs and associated standard errors for college sports. CFB represents college football, while CBB-M and CBB-W represent men's and women's college basketball, respectively. CBB-W uses the corrected positive semidefinite fit, whose covariance matrix is on a rank-one boundary; its confidence intervals and fixed effect tests use the Kenward--Roger approximation described in Section~\ref{sec:estimation}. A dash indicates that the covariance-component test has not been recalibrated for this corrected fit.}
\label{tab:tabone}
\vspace{0.5em}
\begin{tabular}{lrrr}
  \toprule
   &CFB&CBB-M&CBB-W\\

  \midrule
\multicolumn{3}{l}{\textit{Estimate}}\\
$\hat{\alpha}_0$ & 2.352 &  2.969 &  2.817  \\ 
  $\hat{\alpha}_1$ & -0.102 &  -0.074 & -0.0485  \\
  $\hat{\sigma}^2_{\lambda}$&1.020&1.100&1.048\\
  $\hat{\sigma}^2_{1}$&0.946&0.503&0.364 \\
  $\hat{\sigma}_{12}$&0.106&0.001&-0.0022\\
  $\hat{\sigma}^2_{2}$&0.013&0.001&1.4e-5\\
    mean$(w^2_{2017\cdot})$&4.090&1.144&1.289 \\
     \midrule
\multicolumn{3}{l}{\textit{95\% Confidence Intervals}}\\
$\alpha_0$ lower&1.402&2.661&2.535 \\
$\alpha_0$ upper&3.302&3.276&3.099 \\
$\alpha_1$ lower&-0.205&-0.097&-0.069 \\
$\alpha_1$ upper&0.002&-0.051&-0.028\\
   \midrule
\multicolumn{3}{l}{\textit{p-values for null hypothesis}}\\
$\alpha_0=0$ & 3e-4  &5e-19   & 1e-17  \\
  $\alpha_1=0$ & 0.053 & 3e-7 &8e-5\\
 $\bsG=\bds{0}$ & 0.169&$<$1e-5&---\\
   \bottomrule	
\end{tabular}
\end{table}

\begin{table}
\caption{Phase II results of fitting the linear model (\ref{eq:fullfixed}) to the Phase I estimated HFAs and associated standard errors for professional sports. p-values at bottom are generated by likelihood ratio tests of model (\ref{eq:fullfixed}) against reduced models (\ref{eq:fixed1}) and (\ref{eq:fixed2}), respectively  }
\label{tab:pro}
\vspace{0.5em}
\begin{tabular}{lrrr}
  \toprule
 & NFL & NBA & WNBA \\ 
  \midrule
\multicolumn{3}{l}{\textit{Estimate}}\\
$\hat{\beta}_{0A}$ & 2.753 & 3.333 & 3.705 \\ 
  $\hat{\beta}_{0B}$ & -0.276 & -0.775 & -1.702 \\ 
  $\hat{\beta}_{1A}$ & 0.035 & -0.021 & -0.013 \\ 
  $\hat{\beta}_{1B}$ & -0.036 & -0.023 & -0.075 \\ 
  $\sigma^2_{\lambda}$ & 0.865 & 1.096 & 0.715 \\ 
  mean$(w^2_{2017\cdot})$ & 1.618 & 0.400 & 2.936 \\
   \midrule
\multicolumn{3}{l}{\textit{95\% Confidence Intervals}}\\
$\hat{\beta}_{0A}$ lower & 1.673 & 2.776 & 2.641 \\ 
  $\hat{\beta}_{0A}$ upper & 3.832 & 3.890 & 4.770 \\ 
  $\hat{\beta}_{0B}$ lower & -1.796 & -1.566 & -3.277 \\ 
  $\hat{\beta}_{0B}$ upper & 1.243 & 0.016 & -0.128 \\ 
  $\hat{\beta}_{1A}$ lower & -0.074 & -0.077 & -0.113 \\ 
  $\hat{\beta}_{1A}$ upper & 0.144 & 0.035 & 0.087 \\ 
  $\hat{\beta}_{1B}$ lower & -0.188 & -0.101 & -0.223 \\ 
  $\hat{\beta}_{1B}$ upper & 0.116 & 0.056 & 0.073 \\ 
   \midrule
\multicolumn{3}{l}{\textit{p-value for null hypothesis}}\\
$\beta_{0B}=\beta_{1B}=0$ & 0.874 & 0.011 & 0.013 \\ 
  $\beta_{1A}=\beta_{1B}=0$ & 0.784 & 0.179 & 0.221 \\ 
   \bottomrule
\end{tabular}
\end{table}

Most estimates of $\sigma^2_{\lambda}$ in Tables~\ref{tab:tabone} and \ref{tab:pro} are near 1 due to the use of the squared reciprocals of the standard errors of $\hat{\lambda}$ as weights in the linear models. Estimates of error variance away from 1 indicate either over- or under-dispersion in the Phase II model from the estimated Phase I variation.

\subsubsection{College Results}
The estimates $\hat{\alpha}_1$ and p-values for the test $\alpha_1=0$ in Table~\ref{tab:tabone} indicate the strength of the trend of the population of conference-level HFAs over time. While college football produces the steepest downward slope of the three college sports, there is not strong evidence that this slope is not an artifact of conference and error variability over the observed period (p=0.053). Individually, the SEC and PAC10/12 conferences show an increase in estimated HFA over this time period, while the steepest decreases occur in the MAC, Sun Belt, and WAC conferences (see the supplementary data).

By contrast, there is a strong indication of downward linear trends in the men's and women's college basketball HFA. With estimates of $-0.074$ (p~=~3e-07) and $-0.0485$ (p~=~8e-05), respectively, these translate into decreases of about 1.26 and 0.82 points between 2000 and 2017, a span of 17 years across 18 observed seasons. The magnitudes of these decreases correspond to about 42\% and 29\% of the estimated 2017 HFA ($\hat{\alpha}_0$) in each sport, making them appear practically as well as statistically significant. We can conclude that there has been a significant and noticeable decrease in the HFA of men's and women's college basketball intraconference games from 2000 to 2017.

There is evidence of conference-level heterogeneity in the HFA trends. Variability in conference-level intercepts in all three sports is evident from Figure~\ref{plot:HFE_b}, which also reveals variability in conference-level slopes in college football. This variability is also reflected in the magnitudes of the $\hat{\sigma}^2_1$ and $\hat{\sigma}^2_2$ estimates in Table~\ref{tab:tabone}.

\begin{figure}
\caption{Scoring HFA for intraconference games for the Ivy League and Big 12 conferences of men's and women's college basketball. }
\label{plot:ivy}
\centering
\includegraphics[scale=0.5]{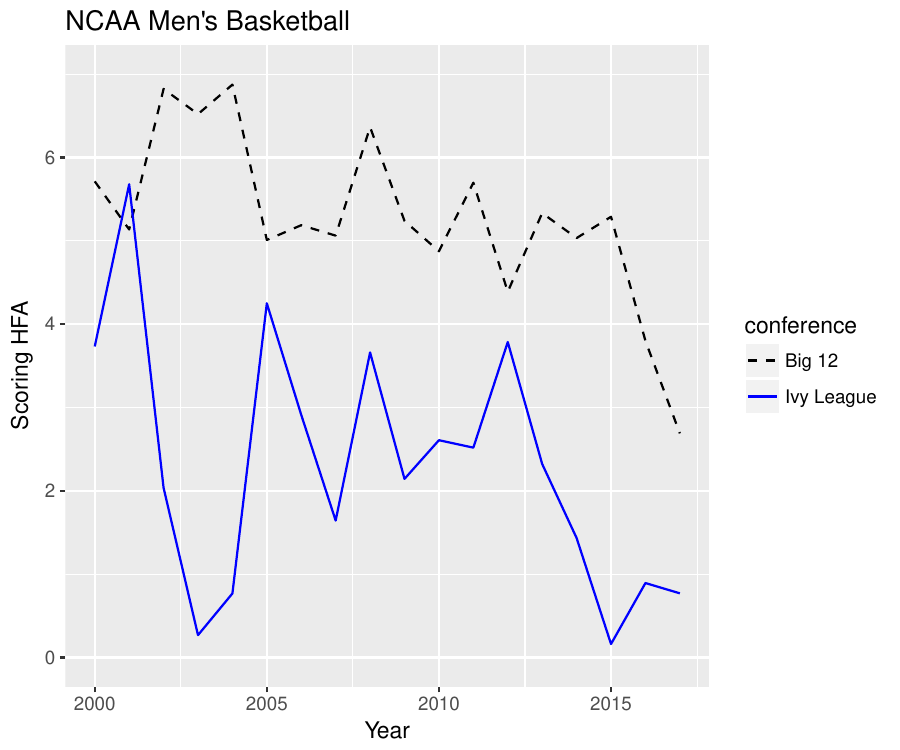}
\includegraphics[scale=0.5]{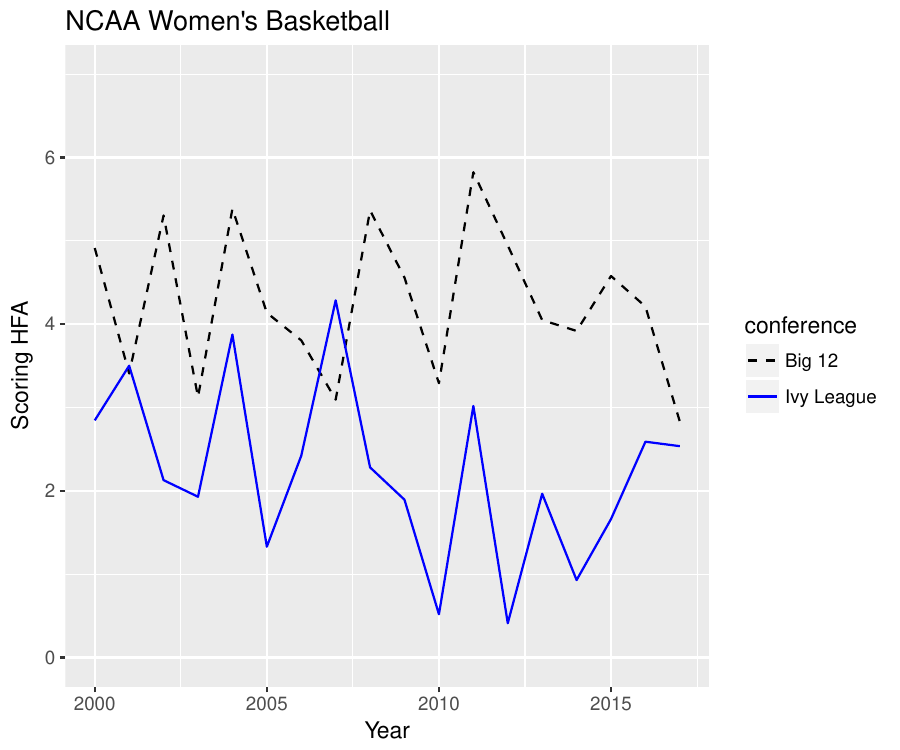}
\end{figure}

While Figure~\ref{plot:HFE_b} does not include conference labels, the source file (available in the supplementary material)  reveals similarity in the conference-level HFA estimates from men's and women's college basketball: in both cases, the Big 12 consistently had among the largest scoring HFAs while the Ivy League had among the smallest (Figure~\ref{plot:ivy}).

\subsubsection{Professional Results}

We test for a linear trend in intraconference HFA in the professional sports with a likelihood ratio test of the full model (\ref{eq:fullfixed}) against the reduced model (\ref{eq:fixed2}) with $\beta_{1A}=\beta_{1B}=0$. The resulting p-values from this test appear in the last row of Table~\ref{tab:pro}: there is no indication of a sustained trend in any of the three professional sports. This is consistent with the plots of the HFAs in the second column of Figure~\ref{plot:HFE_b}.

The NBA and WNBA plots suggest a difference between the trends of the Eastern and Western conference HFAs. In the WNBA, the fitted 2017 Eastern conference HFA is 1.7 points ($1.702/3.705=46\%$) lower than the Western conference HFA, while in the NBA the fitted 2017 Eastern conference HFA is 0.8 points ($0.775/3.333=23\%$) lower than the Western conference HFA. This is formally tested in a likelihood ratio test comparing the full model (\ref{eq:fullfixed}) against the reduced model (\ref{eq:fixed1}) where $\beta_{0B}=\beta_{1B}=0$: the resulting p-values appear in the penultimate row of Table~\ref{tab:pro}. With p-values near 0.01, the tests offer slight (keeping in mind the large number of p-values being reported in this paper) but inconclusive support for the difference in conference slopes and intercepts.

\section{Concluding Remarks}
\label{sec:discussion}
Division I college men's and women's basketball intraconference games showed strong downward linear trends in scoring HFA over the 2000--2017 seasons. While college football shows an even steeper decrease in intraconference HFA over the same period, there is also much more variability in the Phase I within-season estimates in college football than in college basketball. There is also greater conference-level variability in the Phase II estimates of slopes and intercepts in the HFA linear trends for football. Due to the magnitude of the variability, it is difficult to say if the observed downward trend in college football intraconference HFA represents a systematic shift in population behavior or is merely an artifact of a sample from a noisy process with constant mean.

\section*{Appendix: $p$-hacking Disclosure}
The American Statistical Association (ASA) statement on p-values \citep{pvalue} and the associated commentaries call for a reform in the approach and presentation of statistical analyses. Among its recommendations are that journal authors document additional tests and analysis that were performed but not included in the text of the article.

\begin{itemize}
\item{Before committing to the analysis of linear trends in HFA, we examined a plot of the estimates to see that a line provided a plausible explanation for the trend. Had there been evidence of curvature or autocorrelation, the analysis would have likely taken a different approach. We also originally allowed for correlation in the Phase II professional results of model (\ref{eq:fullfixed}) by adding Toeplitz bands to the error covariance matrix, but chose not to include these effects in the model presented in the paper after finding no significant correlation. Note that the random coefficient model (\ref{eq:rancoef}) for the college results naturally incorporates correlation at the conference level.}
\item{The flexibility to choose a starting and ending year for the study comes with a potential danger of data dredging by selecting these time points based on the results of the analysis. The decision to exclude any game results beyond the end of the 2017 season for each sport was made in advance and based on the manuscript timeline, while the decision to begin with the 2000 season was determined by data set availability and a desire to compare all of the sports over the same seasons. Early drafts did consider additional prior seasons for some sports: no major changes to the conclusions would have resulted from inclusion of these seasons, although there would have been a need to address curvature in the football and men's basketball HFA prior to 2001.}
\item{Besides the statistical tests reported here, many others were performed during research for this paper. For example, an early draft only considered full season results from the Phase I mixed model, and also explored trends in the estimated home and away mean scores and conditional home and away error variances from a joint model for home and away scores \citep[Section 2.1]{broatch3}. These were abandoned for the sake of providing a tighter narrative; however, these fitted values remain in the full season Phase I results of the supplementary data and may be further explored by the reader. Even though the original intention of the research was to examine the strength of linear trends in HFA, the number of such other p-values considered during manuscript preparation must be taken into account when considering the strength of the p-values reported in the paper.}
\end{itemize}

\bigskip

\bibliographystyle{agsm}

\bibliography{disbib4}

@Article{broatch3,
  Title                    = {Multivariate Generalized Linear Mixed Models for Joint Estimation of Sporting Outcomes},
  Author                   = {Jennifer E. Broatch and Andrew T. Karl},
  Journal                  = {arXiv:1710.05284},
  Year                     = {2017}
}

@Article{goumas,
  author    = {Chris Goumas},
  title     = {Home advantage and referee bias in {European} football},
  doi       = {10.1080/17461391.2012.686062},
  eprint    = {http://dx.doi.org/10.1080/17461391.2012.686062},
  note      = {PMID: 24444213},
  number    = {sup1},
  pages     = {S243-S249},
  volume    = {14},
  journal   = {European Journal of Sport Science},
  publisher = {Taylor \& Francis},
  year      = {2014},
}

@Article{harville03,
  Title                    = {The Selection or Seeding of College Basketball or Football Teams for Postseason Competition},
  Author                   = {David A. Harville},
  Journal                  = {Journal of the American Statistical Association},
  Year                     = {2003},
  Number                   = {461},
  Pages                    = {17-27},
  Volume                   = {98},
}

@Article{harville77,
  Title                    = {The Use of Linear-Model Methodology to Rate High School or College Football Teams},
  Author                   = {David Harville},
  Journal                  = {Journal of the American Statistical Association},
  Year                     = {1977},
  Number                   = {358},
  Pages                    = {278-289},
  Volume                   = {72},
}

@Article{harville94,
  author    = {David A. Harville and Michael H. Smith},
  title     = {The Home-Court Advantage: How Large is it, and Does it Vary from Team to Team?},
  doi       = {10.1080/00031305.1994.10476013},
  eprint    = {http://www.tandfonline.com/doi/pdf/10.1080/00031305.1994.10476013},
  number    = {1},
  pages     = {22-28},
  volume    = {48},
  journal   = {The American Statistician},
  publisher = {Taylor \& Francis},
  year      = {1994},
}

@Article{henderson75,
  Title                    = {Best Linear Unbiased Estimation and Prediction under a Selection Model},
  Author                   = {C. R. Henderson},
  Journal                  = {Biometrics},
  Year                     = {1975},
  Pages                    = {423--447},
  Volume                   = {31},
  Number                   = {2},
  Doi                      = {10.2307/2529430},
}

@Article{laird82,
  Title                    = {Random-Effects Models for Longitudinal Data},
  Author                   = {N. M. Laird and J. H. Ware},
  Journal                  = {Biometrics},
  Year                     = {1982},
  Pages                    = {963-974},
  Volume                   = {38},
}

@Article{marshall,
  Title                    = {Home Advantage in the {NBA} as a Game-Long Process},
  Author                   = {Marshall B Jones},
  Journal                  = {Journal of Quantitative Analysis in Sports},
  Year                     = {2007},
  Number                   = {4},
  Volume                   = {3}
}

@Misc{masseyr,
  Title                    = {Massey {Ratings}},
  Author                   = {Ken Massey},
  Url                      = {https://www.masseyratings.com/data.php},
  Year                     = {2017},

  Pages                    = {https://www.masseyratings.com/data.php}
}

@Manual{mvglmmRank,
  Title                    = {mvglmmRank: Multivariate Generalized Linear Mixed Models for Ranking Sports Teams},
  Author                   = {Andrew T. Karl and Jennifer Broatch},
  Note                     = {{R} package version 1.0-2},
  Year                     = {2014}
}

@Article{omaha,
  Title                    = {College football's greatest myth},
  Author                   = {Dirk Chatelain},
  Journal                  = {Omaha World-Herald},
  Year                     = {2016},

  Date                     = {2017-09-01},
  Url                      = {http://dataomaha.com/bigstory/sports/college-footballs-greatest-myth}
}

@Article{pollard05,
  author    = {R Pollard and G Pollard},
  title     = {Long-term trends in home advantage in professional team sports in {N}orth {A}merica and {E}ngland},
  doi       = {10.1080/02640410400021559},
  eprint    = {http://dx.doi.org/10.1080/02640410400021559},
  number    = {4},
  pages     = {337-350},
  volume    = {23},
  journal   = {Journal of Sports Sciences},
  publisher = {Routledge},
  year      = {2005},
}

@Article{pvalue,
  author    = {Ronald L. Wasserstein and Nicole A. Lazar},
  title     = {The {ASA}'s Statement on p-Values: Context, Process, and Purpose},
  doi       = {10.1080/00031305.2016.1154108},
  eprint    = {http://dx.doi.org/10.1080/00031305.2016.1154108},
  number    = {2},
  pages     = {129-133},
  volume    = {70},
  journal   = {The American Statistician},
  publisher = {Taylor \& Francis},
  year      = {2016},
}

@TechReport{pxem,
  Title                    = {A faster and computationally more efficient {REML} {(PX)EM} algorithm for linear mixed models},
  Author                   = {Simon Diffey and Alan Welsh and Alison Smith and Brian R. Cullis},
  Institution              = {Centre for Statistical and Survey Methodology, University of Wollongong},
  Year                     = {2013},
  Number                   = {2-13},
  Url                      = {http://ro.uow.edu.au/cssmwp/108}
}

@Article{scheipl,
  Title                    = {Size and power of tests for a zero random effect variance or polynomial regression in additive and linear mixed models},
  Author                   = {Fabian Scheipl and Sonja Greven and Helmut Kuchenhoff},
  Journal                  = {Computational Statistics \& Data Analysis},
  Year                     = {2008},
  Number                   = {7},
  Pages                    = {3283 - 3299},
  Volume                   = {52},

  Doi                      = {https://doi.org/10.1016/j.csda.2007.10.022},
  ISSN                     = {0167-9473},
}

@Article{si2,
  Title                    = {Hoop Thoughts: What is happening to home court advantage?},
  Author                   = {Seth Davis},
  Journal                  = {Sports Illustrated},
  Year                     = {2017},

  Date                     = {2017-01-09},
  Issuetitle               = {Sports Illustrated},
  Organization             = {Sports Illustrated},
  Url                      = {https://www.si.com/college-basketball/2017/01/09/hoop-thoughts-home-court-advantage-baylor-kansas-villanova}
}

@Book{stroup,
  Title                    = {Generalized Linear Mixed Models: Modern Concepts, Methods and Applications},
  Author                   = {Walter W. Stroup},
  Publisher                = {Boca Raton: CRC Press},
  Year                     = {2012},
}

@Article{wang,
  Title                    = {Home Advantage in American College Football Games: A Multilevel Modelling Approach},
  Author                   = {Winnie Wang and Ron Johnston and Kelvyn Jones},
  Journal                  = {Journal of Quantitative Analysis in Sports},
  Year                     = {2011},
  Volume                   = {7},
  Number                   = {3},
  Doi                      = {10.2202/1559-0410.1328}
}

@Article{KarlZimmerman2021,
  author = {Andrew T. Karl and Dale L. Zimmerman},
  title = {A diagnostic for bias in linear mixed model estimators induced by dependence between the random effects and the corresponding model matrix},
  journal = {Journal of Statistical Planning and Inference},
  year = {2021},
  volume = {211},
  pages = {107--118},
  doi = {10.1016/j.jspi.2020.06.004}
}
\end{document}